\documentclass[journal]{IEEEtran}
\usepackage{siunitx} 
\usepackage{multirow}
\usepackage{xcolor}
\definecolor{redcolor}{rgb}{1, 0, 0}
\usepackage{graphicx}
\usepackage{psfrag}  
\usepackage{subfigure}
\usepackage{tabularx}
\usepackage{longtable}
\usepackage{threeparttable}
\usepackage{textcomp}
\usepackage{amsmath, bm}
\usepackage[font=footnotesize]{caption}
\usepackage{cite}

\begin{document}
\bibliographystyle{IEEEtran}
\nocite{*}
\title{Low Phase Noise Oscillator Topology for mm-Wave Voltage-Driven Current-Sensed High-Q Active On-Chip MEMS Resonator} 
\title{Low Phase Noise Oscillator for mm-Wave Voltage-Driven Current-Sensed High-Q Active-Mode On-Chip MEMS Resonator} 
\title{mmWave Oscillator Topology for Improving Fundamental Limits of Phase Noise using a High-Q Active-Mode On-Chip MEMS Resonator}
\title{mmWave Oscillator Topology with Improved Fundamental Limits of Phase Noise using a High-Q Active-Mode On-Chip MEMS Resonator}
\title{A mmWave Oscillator Design Utilizing High-Q Active-Mode On-Chip MEMS Resonators for Improved Fundamental Limits of Phase Noise}
\author{Abhishek~Srivastava,~\IEEEmembership{Senior Member,~IEEE,}
        Baibhab~Chatterjee,,~\IEEEmembership{Graduate Student Member,~IEEE,}
        Udit~Rawat,~\IEEEmembership{Graduate Student Member,~IEEE,}
        Yanbo~He,~\IEEEmembership{Member,~IEEE,}
        Dana~Weinstein,~\IEEEmembership{Senior Member,~IEEE,}
        and~Shreyas~Sen,~\IEEEmembership{Senior Member,~IEEE}
}

\maketitle
\begin{abstract}
Recent progress in MEMS resonant-fin-transistors (RFT) allows very high-Q active mode resonators, promising crystal-less monolithic clock generation for mmWave systems. 
However, there is a strong need for design of mmWave oscillators that utilize the high-Q of active-mode RFT (AM-RFT) optimally, while handling unique challenges such as 
resonator's low electromechanical transduction. 
In this brief, we develop a theory and through design and post-layout simulations in 14-nm Global Foundry process, 
we show the first active oscillator with AM-RFT 
at 30 GHz, which 
improves the fundamental limits of phase noise and figure-of-merit 
as compared to the oscillators with conventional LC resonators. 
For AM-RFT %
with Q-factor of 10K, post layout simulation results show that the proposed oscillator exhibits phase noise $<$-140 dBc/Hz and figure-of-merit $>$228 dBc/Hz at 1 MHz offset for 30 GHz center frequency, which are $>$25 dB better than the existing monolithic LC oscillators.
\end{abstract}
\renewcommand\IEEEkeywordsname{Keywords}
\begin{IEEEkeywords}
Oscillator, mmWave, MEMS, phase-noise, 14 nm
\end{IEEEkeywords}

\section{Introduction}
Utilization of mmWave spectrum ($>$20 GHz) in 5G communication technologies for increased data rates (upto 4 Gb/s) requires large arrays of wireless transceivers in multiple-input-multiple-output (MIMO) fashion \cite{mimo_2016}. For cost effective and area efficient mmWave carrier generation in MIMO systems, it requires more versatile resonators with monolithic IC integration capabilities  
at mmWave frequencies 
as compared to the 
conventional phase locked loop (PLL) based carrier synthesis, which needs bulky and costly off-chip piezoelectric quartz crystals. 
With the advancements in micro-electro-mechanical systems (MEMS) technologies, 
such as MEMS-First, MEMS-Last \cite{mems_first_last}, Back-End-Of-Line MEMS \cite{mems_beol} and Acoustic Bragg Reflectors (ABR) \cite{nawell_abr},  
it seems viable to satisfy the demands of fully monolithic mmWave carrier synthesis. 
Traditional MEMS resonators utilize passive electrostatic or
piezoelectric transduction schemes, and exhibit low electromechanical transduction and high parasitic feed-through, 
which can be mitigated by active FET sensing techniques \cite{reso_electrical_model_11GHz_2013}. 
In our previous work \cite{rft_isscc2018}, we utilized active FET sensing and demonstrated 
a very high-Q, active mode 
Resonant Fin Transistor (AM-RFT) using ABR technique. %
AM-RFT was fabricated in Global Foundary's 14nm FinFET technology, leveraging the vertical 3D geometry of FinFETs to efficiently confine, drive, and sense acoustic vibrations in the solid (unreleased) CMOS stack. 
In this work, we utilize high-Q active-mode MEMS resonator to develop an oscillator %
for fully monolithic mmWave carrier synthesis, for which 
\begin{figure}
\begin{center}
\includegraphics[scale=0.5]{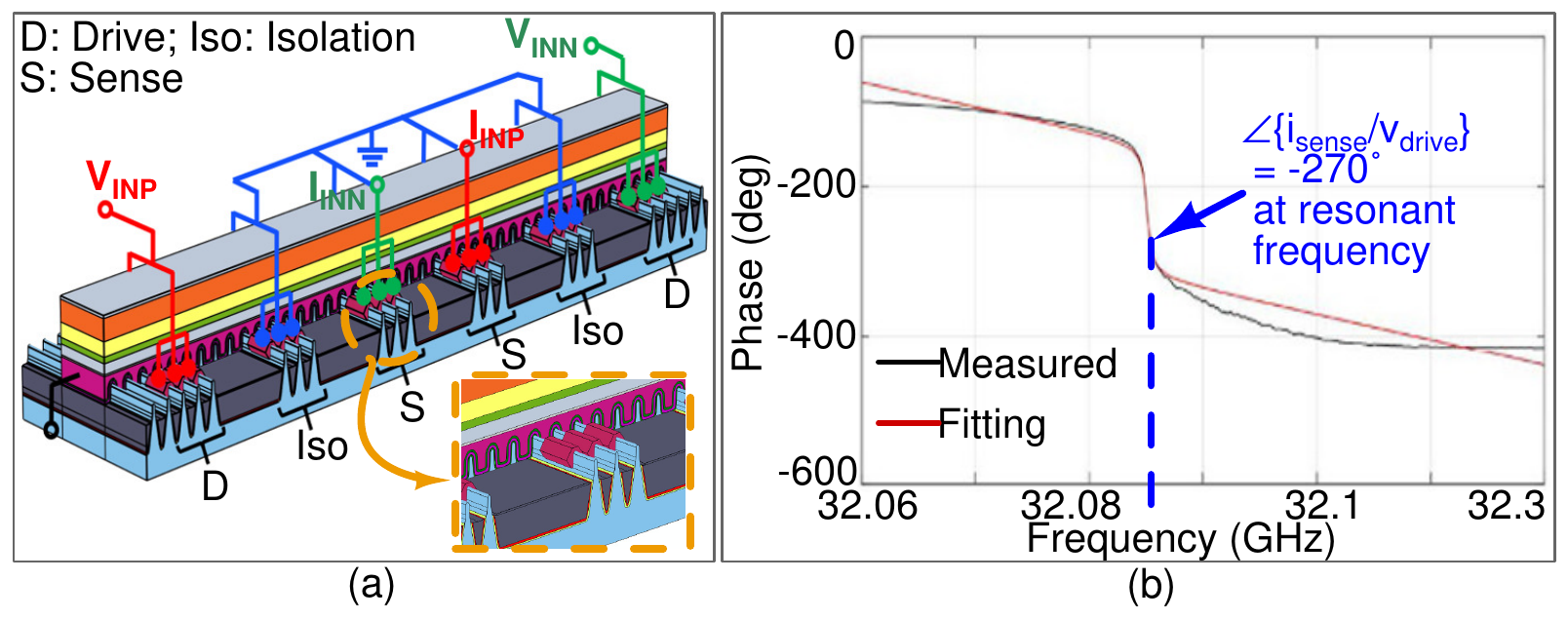}
\vspace{-2mm}
\caption{(a) RFT unit cell showing differential driving voltage for a 3-skip-4 fin connections, (b) measured phase shift between output (sensed) current and input (drive) voltage \cite{rft_isscc2018}}
\label{fig_rft_fin}
\end{center}
\vspace{-8mm}
\end{figure}
we present - 
1) a novel oscillator topology with active-mode MEMS resonator, 
which improves the existing fundamental limits of phase noise and figure-of-merit (FoM) obtained from LC oscillators,  
2) an electrical model of voltage-driven, current-sensed high-Q on-chip active-mode MEMS resonator for circuit simulation, 
3) an analysis for phase noise of the proposed oscillator
and 
4) verification of the proposed topology with the design and simulation of a 30 GHz oscillator in 14-nm GF process. 
In this brief, AM-RFT has been used to explain the techniques proposed, however, the insights presented can be in general applied for any high-Q on-chip active-mode MEMS resonator. 

The paper is organised as follows. Section \ref{mems} presents the device structure and electrical model of AM-RFT. 
Sections \ref{ckt_topo} and \ref{sec_phase_noise} present the proposed oscillator topology and its phase noise analysis, respectively. 
Section \ref{impl_result} presents the implementation details with post-layout simulation results followed by the conclusion of the paper. 
\section{Active Mode MEMS Resonator}
\label{mems}
\subsection{Device structure}
The central section of the complete AM-RFT cavity is shown in Fig.\ref{fig_rft_fin}(a). Multiple pairs of drive MOSCAP transducer units are used for converting the input voltage capacitively into the actuation force. Each unit consists of a 3-fin MOSCAP with the Gate forming one of the plates and the shorted Source-Drain forming the other. The respective phases of the differential signal are applied to the shorted Source-Drains of the the drive units. To satisfy the DRC constraints 4 fins are skipped between every pair of units. Once the resonance modes are excited by applying the differential input actuation voltage, the vibrating fins exhibit periodically varying mechanical stress. The periodic stress modulates the drain current flowing through the 3-fin sense transistors resulting in a differential output (due to opposite phases of stress at the sense transistor fins). 
A Phase shift of 90\textdegree\ results for the transformation from electrical to mechanical domain. Subsequently, the output current of the sense transistors in conjunction with their output resistances causes an additional phase shift of 180\textdegree\,. 
Essentially, ABR based active mode resonators act as a voltage controlled current source with 270\textdegree\ phase shift (VCCS$_{270}$) between the input (driving) voltage and the output (sensed) drain current. 
Fig.\ref{fig_rft_fin}(b) also shows that the measured phase shift for AM-RFT is 270\textdegree\ \cite{rft_isscc2018}.

\vspace{-3mm}
\subsection{Electrical Model for Active Mode RFT}
\begin{figure}%
\centering
\includegraphics[scale=0.39]{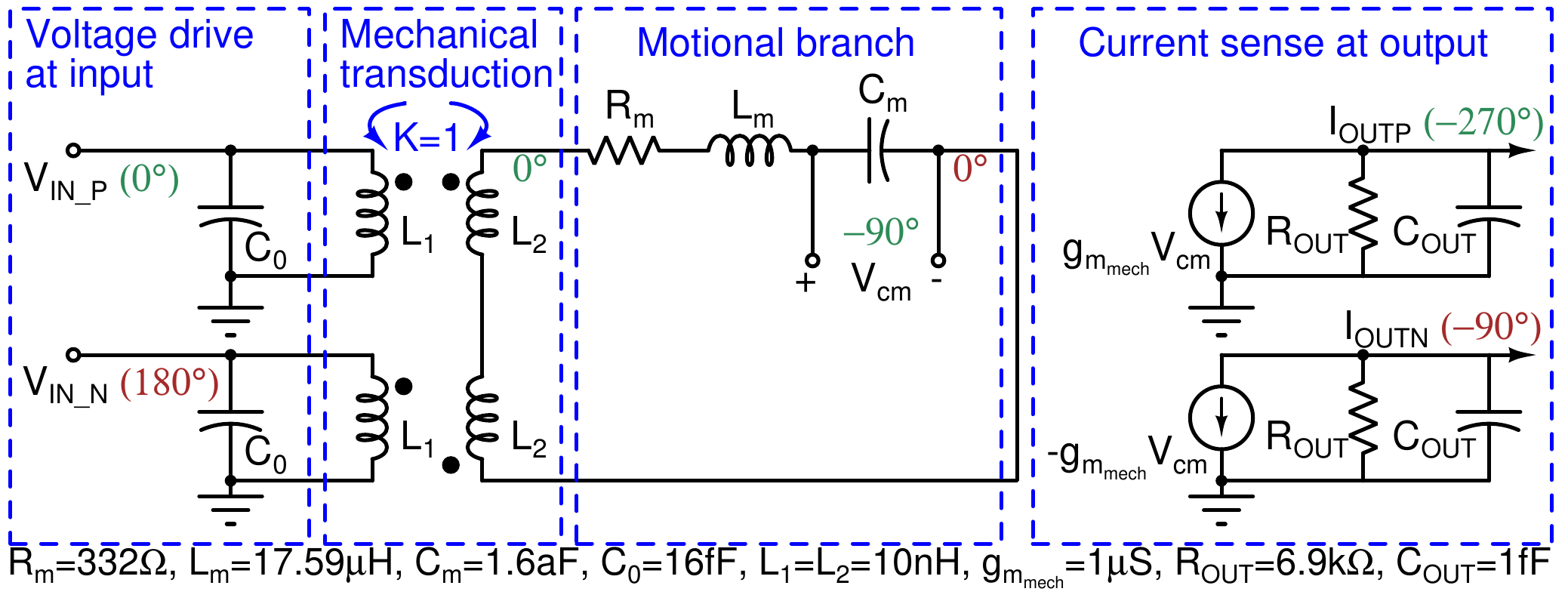}%
\vspace{-2mm}
\caption{Proposed electrical model of fully differential active mode RFT for circuit simulations}
\label{fig_reso}
\vspace{-6mm}
\end{figure}
Fig. \ref{fig_reso} depicts the proposed electrical model %
for fully differential VCCS$_{270}$ realization of AM-RFT. %
$R_m$, $L_m$ and $C_m$ are the motional resistance, inductance and capacitance, respectively, which model the resonant behaviour of the device with resonant frequency $f_0 = \frac{1}{2\pi \sqrt{L_mC_m}}$ \cite{motional_param_2013}, \cite{motional_param_2006}. 
$C_0$ is the static capacitance at the driving ports \cite{rft_isscc2018} and $g_{m_{mech}}$ is the transconduction showing relation between sensed drain current ($I_{OUT}$) and voltage across $C_m$ ($V_{cm}$). 
$R_{OUT}$ and $C_{OUT}$ are the output resistance and capacitance at the drain terminal of the sensing FET, respectively. 
Fig. \ref{fig_reso} also shows the parameters of the proposed electrical model of 30 GHz AM-RFT with Q = 10,000 \cite{rft_isscc2018}, 
where $C_0$, $g_{m_{mech}}$, $R_{OUT}$ and $C_{OUT}$ were evaluated with the help of AC simulations of the layout extracted netlist of the RFT device. 
For evaluating $C_m$, sufficiently small electromechanical coupling coefficient ($\frac{C_m}{C_0}$) of $10^{-4}$ was considered for AM-RFT. 
In AM-RFT, the motional branch directly appears at the driving port due to the mechanical transduction \cite{reso_electrical_model_11GHz_2013}. Therefore, 
as shown in Fig. \ref{fig_reso}, mechanical transduction is modelled with mutually coupled inductors $L_1$ and $L_2$ having coupling co-efficient ($K$) as 1, because any other value of $K$ will 
modify the impedances of the driving port and the motional branch due to the transformer action 
between $L_1$ and $L_2$. 
It is also important to select $L_1=L_2<<L_m$ such that $f_0$ is not changed and voltage drop across $L_1$ and $L_2$ is negligible as compared the drop across the motional branch. At the same time $L_1$ and $L_2$ should be large enough such that they do not load the drive input. Considering these facts, $L_1=L_2 =10$ nH is chosen as shown in Fig. \ref{fig_reso}. 

The model shown in Fig. \ref{fig_reso} helps in topological evolution and simulations of oscillator circuits for active-mode resonators. 
Now two questions arise - 1) how can an active-mode resonator, realised as VCCS$_{270}$, 
be utilized to build an oscillator and 2) 
will there be any advantage in the oscillator phase noise as compared to the conventional LC oscillators. 
These questions are addressed in sections \ref{ckt_topo} and \ref{sec_phase_noise}, respectively.

\section{Oscillator Circuit Topology}
\label{ckt_topo}
\begin{figure}
\centering 
\includegraphics[scale=0.39]{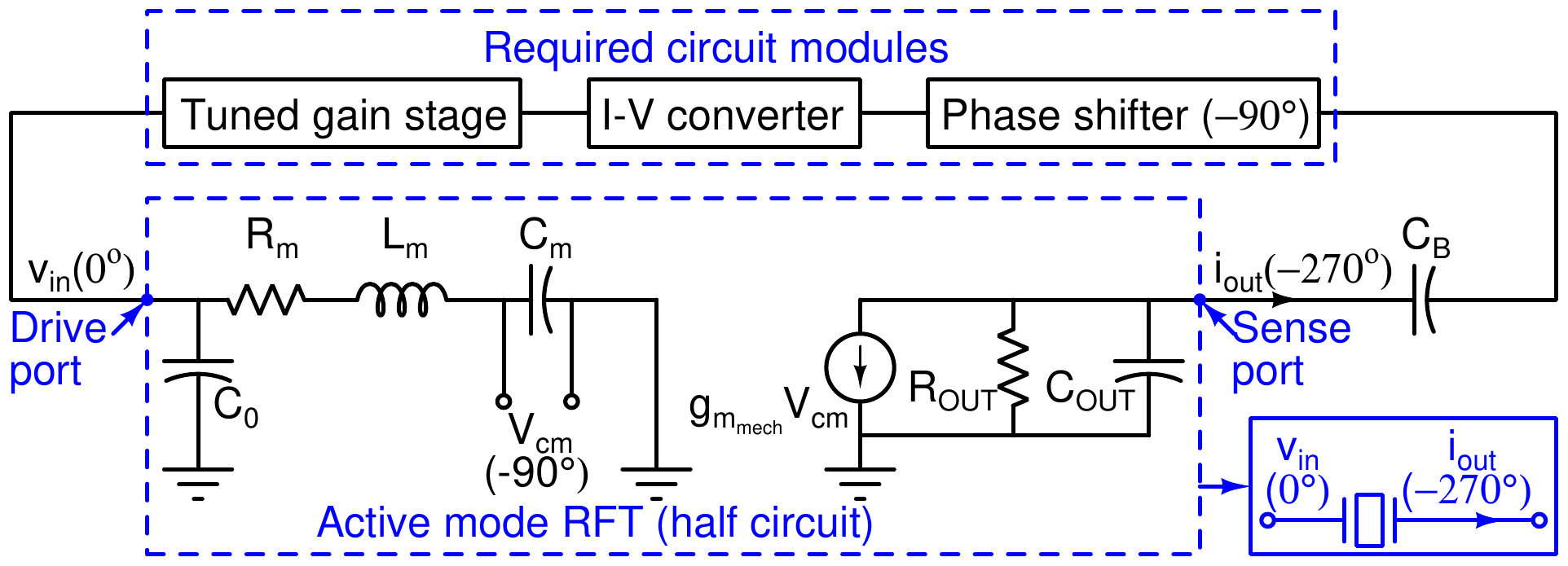}
\vspace{-2mm}
\caption{Depiction of required circuit modules to build oscillator with half circuit model of VCCS$_{270}$ realization of AM-RFT in a feedback loop}
\label{fig_topo}
\vspace{-4mm}
\end{figure}
To build an oscillator with VCCS$_{270}$ realization of the resonator,  
Barkhausen criteria must be met in a feedback loop, for which the phase shift around the closed loop should be 0\textdegree\ and the loop gain must be $\geq$1. 
Fig. \ref{fig_topo} depicts the oscillator block diagram 
with the help of the half circuit model of VCCS$_{270}$. 
As shown in Fig. \ref{fig_topo}, 
the circuit requires a phase shifter for the additional $-90$\textdegree\ phase shift and an I-V conversion stage 
to convert the current sensed from the output of VCCS$_{270}$ into voltage, which can be fed back to its input drive port.  
Moreover, as shown in Fig. \ref{fig_topo}, for meeting the loop-gain criteria at 30 GHz, a tuned narrowband gain stage is also needed in the loop. 
Realization techniques for phase-shifter and I-V converter, and the order in which the required blocks are placed in the loop lead to different topological choices, which are discussed in the following subsection. 
\subsection{Topology choices} 
\begin{figure}
\centering 
\includegraphics[scale=0.41]{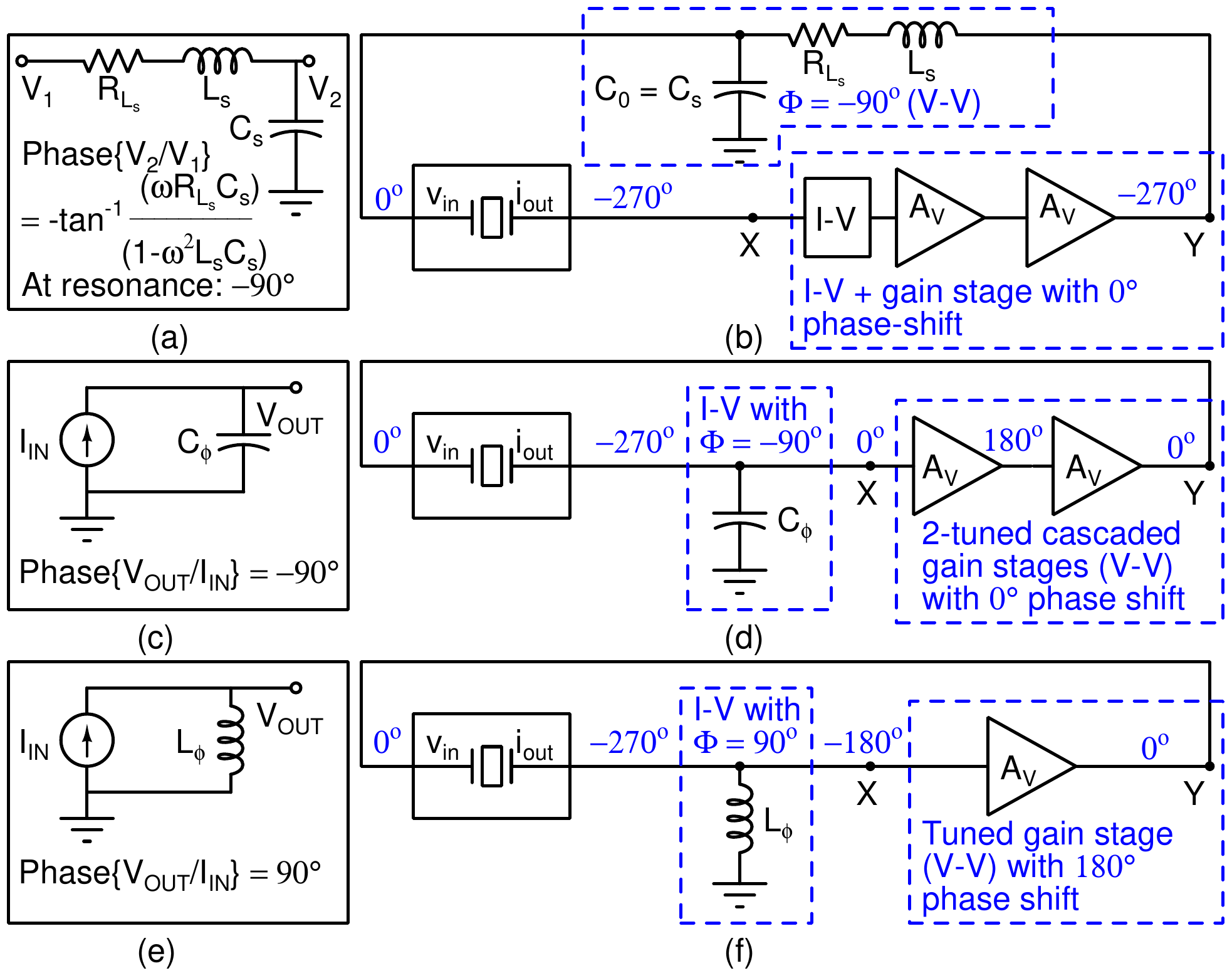}
\caption{Methods to provide 90\textdegree\ phase shift and corresponding oscillator topology choices. (a)-(b) using a series RLC as phase shifter, (c)-(d) using a capacitor as phase shifter and (e)-(f) using an inductor as phase shifter} 
\label{fig_topo_choices}
\vspace{-4mm}
\end{figure}

Fig. \ref{fig_topo_choices} presents three techniques to achieve additional 90\textdegree\ phase shift 
and the corresponding oscillator topologies, which are discussed below. 
\paragraph{Gain-stage-first with a series RLC phase shifter} 
Fig. \ref{fig_topo_choices}(a) depicts the first method, 
where $-$90\textdegree\ phase shift can be obtained at resonance by using a series $R_sL_sC_s$ network. 
Fig. \ref{fig_topo_choices}(b) depicts the oscillator topology with a gain-stage-first approach 
with the series $R_sL_sC_s$ network as phase-shifter, 
where $C_0$ of the resonator is utilized as $C_s$. 
This topology requires a cascade of two tuned gain stages,  
to provide high gain at 30 GHz, along with an active I-V converter 
in order to maintain the phase criteria in the loop, which considerably increases power, area and circuit complexity. 
Therefore, it is not a preferable choice. 
\paragraph{Capacitor as a phase shifter and I-V converter} Fig. \ref{fig_topo_choices}(c) shows the second method, where a capacitor ($C_{\phi}$) is used to provide the phase shift of $-90$\textdegree\, while also doing the I-V conversion. 
Fig. \ref{fig_topo_choices} (d) depicts the topology with $C_{\phi}$
followed by cascade of two tuned gain stages. 
As compared to the first topology shown in Fig. \ref{fig_topo_choices}(b), it does not need an explicit I-V converter and hence is a better choice. 
However, it also needs cascaded tuned stages, which demands for high power consumption, therefore, a lower power alternate must be explored, which is the third method discussed next. 
\paragraph{Inductor as a phase shifter and I-V converter}
Fig. \ref{fig_topo_choices}(e) shows the third method where an inductor is used to provide 90\textdegree\ phase-shift and I-V conversion simultaneously. 
Fig. \ref{fig_topo_choices}(f) depicts the oscillator topology for this technique. 
To meet the phase shift and gain criteria in the loop, this topology needs a single tuned gain stage. 
It is a lower power choice as compared to $C_{\phi}$ method shown in Fig. \ref{fig_topo_choices}(d) as it requires less stages. 
Therefore, this topology has been chosen to build the oscillator with high-Q on-chip active mode MEMS resonator, which is discussed in detail in the following subsection. 
\subsection{Proposed oscillator topology}%
\begin{figure}
\centering
\includegraphics[scale=0.41]{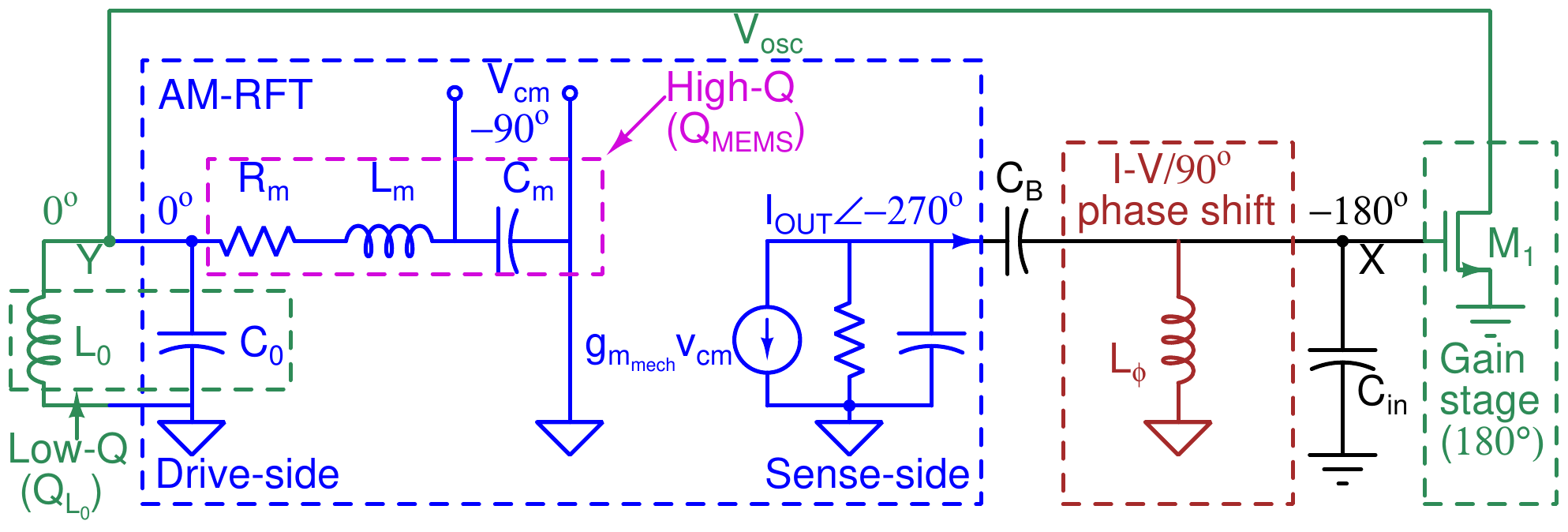}
\caption{Schematic of the proposed oscillator circuit}%
\label{fig_ckt}
\end{figure}

Fig. \ref{fig_ckt} shows the schematic of the proposed oscillator topology, where  
$L_{\phi}$ provides the desired 90\textdegree\ phase shift and I-V conversion. 
For the tuned gain stage, a common source stage is used where the static capacitance ($C_0$) of AM-RFT has been utilized in its load tank with an inductor $L_0$ having a loss resistance $R_{L_0}$ and Q-factor $Q_{L_0} = \frac{\omega L_0}{R_{L_0}}$. 
At mmWave frequencies, the impedance of $C_0$ ($X_{C_0} = \frac{1}{j\omega C_0}$) becomes comparable to $R_m$ and ac signal flows to the ac ground through it, which considerably increases the drive requirements of the oscillator. 
This is a critical problem with on-chip MEMS resonators  \cite{abhishek_tcas2020}, 
which is solved in the proposed topology by including $C_0$ in the resonant tank of the tuned gain stage. 

It is important that at $f_0$, the phase-shifter $L_{\phi}$ does not resonate 
with $C_{in}$, which is the total capacitance due to the routing parasitics and the input of the gain stage $M_1$  (Fig. \ref{fig_ckt}), 
otherwise it will cease to provide 90\textdegree\ phase shift. 
In fact, the resonant frequency ($f_{\phi} = \frac{1}{2\pi \sqrt{L_{\phi}C_{in}}}$) of the $L_{\phi}C_{in}$ tank should be much larger than $f_0$, 
which gives the maximum limit of $L_{\phi}$ as shown in (\ref{eq_l_phi_max}). 
\begin{equation}
\label{eq_l_phi_max}
\small
\frac{1}{2\pi \sqrt{L_{\phi} C_{in}}} >> f_0 
\implies  
L_{\phi} << \frac{1}{4\pi ^2 f_0 ^2C_{in}}
\end{equation}
From Fig. \ref{fig_ckt}, %
gain of tuned stage $A_{V1} = \frac{v_Y}{v_X}$ can be given by (\ref{eq_vy_vx}), while its load tank is tuned near $f_0$.
\begin{equation}
\label{eq_vy_vx}
\small
A_{V1} = \frac	{v_Y}{v_X} = (g_{m_{M1}}\times R_0)^2 
\end{equation}
In (\ref{eq_vy_vx}), $g_{m_{M1}}$ is the transconductance of $M_1$, $R_0 = R_m || (Q_{L_0}^2 \times R_{L_0})|| r_{o_{M1}}$ and $r_{o_{M1}}$ is the small signal output resistance of $M_1$. 
From Fig. \ref{fig_ckt}, $v_X \approx g_{m_{mech}} v_{cm} \times |\omega _0L_{\phi}| = g_{m_{mech}} \frac{v_Y}{R_m} \times \frac{1}{\omega C_m} \times|\omega _0 L_{\phi}|$, which gives 
$\small A_{V2}=\frac{v_X}{v_Y}$ shown in Eq. (\ref{eq_vx_vy}). 
\begin{equation}
\label{eq_vx_vy}
\small
 A_{V2} = \frac{v_X}{v_Y} =  \frac{g_{m_{mech}} L_{\phi}}{C_m {R_m}} 
\end{equation}
For oscillations to build up at $f_0$, $A_{V1}\times A_{V2}  > 1$ $\implies (g_{m_{M1}}\times R_0)^2  \times \frac{g_{m_{mech}} L_{\phi}}{C_m {R_m} }> 1$ , which gives expression for minimum $L_{\phi}$ (\ref{eq_l_phi_min}). 

\begin{equation}
\label{eq_l_phi_min}
L_{\phi} >> \frac{R_m C_m}{g_{m_{mech}}(g_{m_{M1}}\times R_0)^2}
\end{equation}
Equations (\ref{eq_l_phi_max}) and (\ref{eq_l_phi_min}) together suggests the suitable start value for $L_{\phi}$, which can be further optimized through simulations. 

As shown in Fig. \ref{fig_ckt}, there are two resonant tanks in the proposed circuit- 1) High-Q ($Q_{MEMS}$=10K) $R_mL_mC_m$ tank and 2) Low-Q ($Q_{L_0}<$30) $L_0C_0$ tank.
With two tanks in the loop, 
a fundamental question arises for the proposed topology - 
is there any improvement in the phase noise %
as compared to the conventional LC oscillators.
To answer this question, 
phase noise %
analysis for the proposed oscillator is discussed in the following section.
\section{Phase noise performance of the proposed AM-RFT based oscillator} 
\label{sec_phase_noise}
\begin{figure}%
\centering
\includegraphics[scale=0.41]{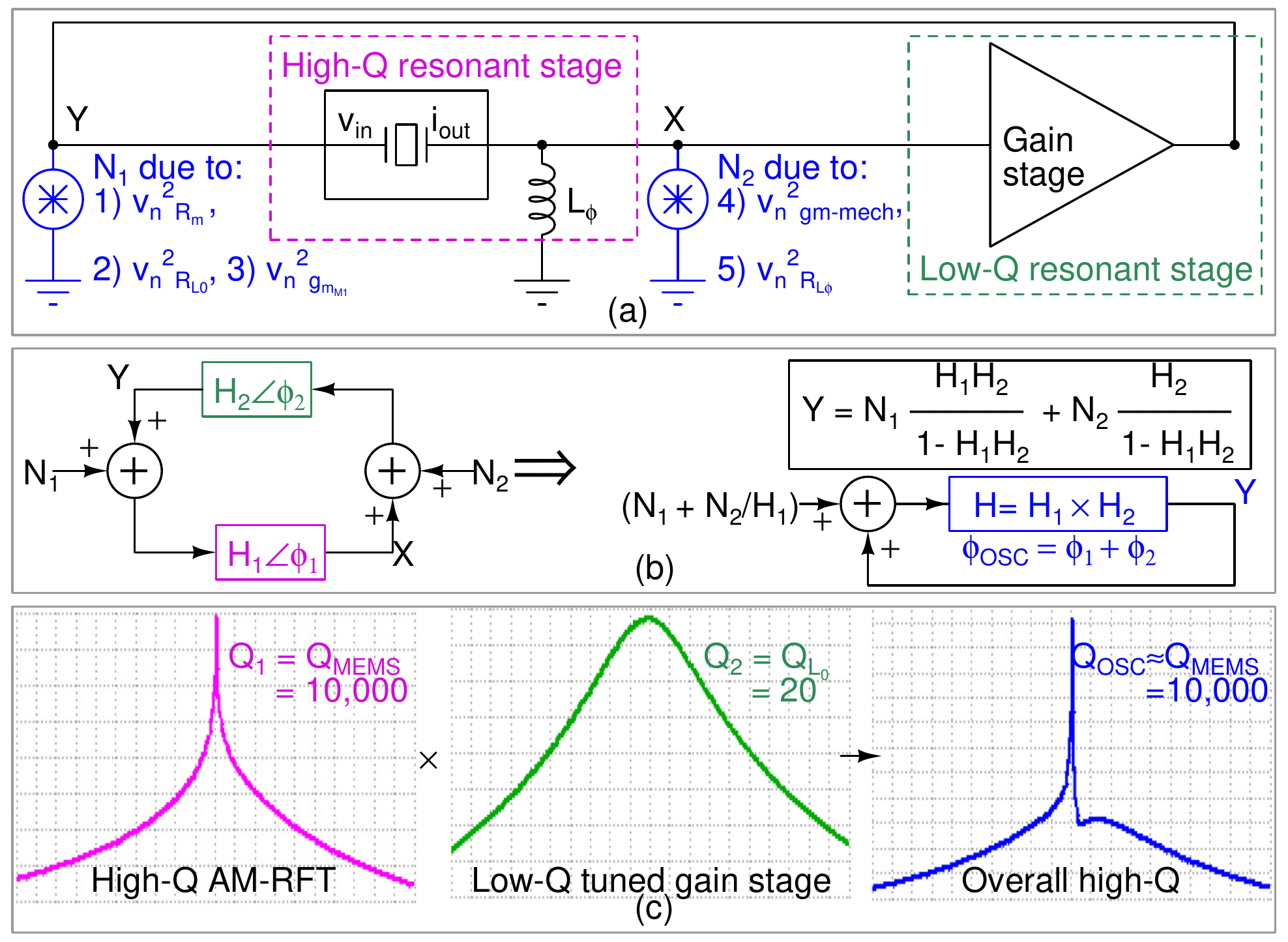}
\caption{(a) Noise sources in the oscillator, (b) depiction of two resonant stages in the positive feedback loop of the proposed oscillator (c) simulation showing overall Q-factor of the oscillator is approximately same as the $Q_{MEMS}$}
\label{fig_pn_improve}
\end{figure}

Phase noise ($\mathcal{L}\{\Delta \omega\}$) in dBc/Hz of an oscillator at an offset ($\Delta\omega$) from a frequency $\omega _{0}$ can be defined as follows.
\begin{equation}
\label{eq_pn}
\small
\mathcal{L}\{\Delta \omega\}  = 10\ log \frac{P_{n}(\Delta\omega)}{P_{c}}\  (dBc/Hz)
\end{equation}
where, $P_{n}(\Delta\omega)$ is the noise power spectral density at $\Delta\omega$ and $P_c = \frac{V_{OSC}^2}{2}$ is the carrier power at $\omega _{0}$ with oscillation amplitude $V_{OSC}$. 
The noise spectrum experiences filtering due to the two resonant circuits present in the loop (Fig. \ref{fig_ckt}), which governs the phase noise performance of the proposed oscillator. 
For modelling the phase noise, effective noise power and its filtering due to the overall Q-factor of the oscillator ($Q_{OSC}$) are discussed below. 
\paragraph{Noise power ($P_n$) sources in the loop}
As shown in Fig. \ref{fig_pn_improve}(a), there are five major noise sources in the oscillator: 
1) $V_{n_{R_m}}^2$ due to $R_m$, 
2) $V_{n_{R_{L_0}}}^2$ due $R_{L_0}$, 
3) $V_{n_{g_{m_{M1}}}}^2$ due to the gain stage's output noise
4) $V_{n_{g_{m_{mech}}}}^2$ due to the sense transistor's output noise and 
5) $V_{n_{R_{L_{\phi}}}}^2$ due to $R_{L_{\phi}}$, 
which are lumped into two noise sources $N_1$ and $N_2$ at nodes X and Y, respectively. 
Noise from $R_{L_0}$ and transistor's channel noise appear directly at the oscillator output. Moreover, impedance reflection (1:1) at input of AM-RFT due to the transformer action, also causes $R_m$ noise to appear at the oscillator output. Gain stage amplifies the total noise present at its input, which is due to the sense transistor's channel noise and thermal noise of $R_{L_{\phi}}$. The total noise power at the oscillator output can be given by Eq. (\ref{eq_pn_all}). 
\begin{multline}
\label{eq_pn_all}
\small
P_n = \bigg(\frac{4kT}{R_m} +\frac{4kT}{R_{L_0}} + 4kT\gamma g_{m_{M1}} \bigg) R_0^2 \\
+ A_{v_1}^2\bigg(4kT\gamma g_{m_{mech}} + \frac{4kT}{R_{L_{\phi}}} \bigg) R_{L_{\phi}}^2 
\end{multline}
\paragraph{Overall Q-factor and effective noise power}
As shown in the Fig. \ref{fig_pn_improve}(b), the two LC resonant circuits are cascaded, therefore their frequency responses get multiplied in the loop ($H(s) = H_1(s) \times H_2(s)$) resulting into an overall phase shift ($\phi	_{OSC}$) given by Eq. (\ref{eq_phase_all}), which is the sum of phase shifts of the two networks.
\begin{equation}
\label{eq_phase_all}
\small
\phi	_{OSC}= \phi _1 + \phi _2
\end{equation}
Since $Q = \frac{\omega _0}{2} |\frac{d\phi}{d\omega}|$ \cite{razavi_phase_noise_1996}, 
therefore, from Eq. (\ref{eq_phase_all}) the overall Q of the oscillator ($Q_{OSC}$) can be given by Eq. (\ref{eq_q_all}). 
\begin{equation}
\label{eq_q_all}
\small
Q_{OSC} = Q_{L_0} + Q_{MEMS} \approx Q_{MEMS}
\end{equation}
Fig. \ref{fig_pn_improve}(c) depicts the simulated frequency responses of the two resonant tanks and their combined  response, which also shows that $Q_{OSC} \approx Q_{MEMS}$ near resonant frequency. 
It is important to note that the 
circuit has two distinct resonant circuits,
which might have slight mismatch in their resonant frequencies due to parasitics and fabrication process. 
However, coupling of the mechanical vibrations through high-Q ($Q_{MEMS}>>Q_{L_0}$) AM-RFT at $f_0$ will result into the  
sustained oscillations closer to $f_0$ with $Q_{OSC} \approx Q_{MEMS}$. 
For the feedback system shown in Fig. \ref{fig_pn_improve}(b), $\frac{Y}{N} = \frac{H}{1-H}$, 
with very high $Q_{OSC}$ and $|H| \approx 1$ near $\omega _0$, which gives $|\frac{d\phi}{d\omega}|^2 >> |\frac{dH}{d\omega}|^2$ leading to Leeson's heuristic expression $|\frac{Y}{N}|^2 = (\frac{\omega_0}{2Q_{OSC}\Delta\omega})^2$ for the proposed topology \cite{leeson_pn}.
Using Leeson's expression, %
$P_{n}(\Delta\omega)$ at an offset of $\Delta\omega$ from $\omega_0$ 
can be given as follows. 
\begin{equation}
\label{eq_leeson}
\small
P_{n}(\Delta\omega) = \bigg(\frac{\omega_0}{2Q_{OSC}\Delta\omega}\bigg)^2P_n
\end{equation}
From Eq. (\ref{eq_pn}) and Eq. (\ref{eq_leeson}), due to very high $Q_{OSC}$, noise filtering is maximized near  resonant frequency, which considerably improves the phase noise of the oscillator. %

\paragraph{Phase noise expression}
Noise factor of oscillator (F), which is defined as the total oscillator phase noise normalized to phase noise due to the MEMS resonator loss ($R_m$), can be calculated 
by dividing (\ref{eq_pn_all}) by $\frac{4kT}{R_m}R_0^2$ and shown in (\ref{eq_noise_fac}).
\begin{multline}
\label{eq_noise_fac}
\small
F = 1 + \frac{R_m}{R_{L_0}} + \gamma g_{m_{M1}} R_m + \gamma g_{m_{mech}}g_{m_{M1}}^2R_mR_{L_{\phi}}^2 \\
 + g_{m_{M1}}^2R_mR_{L_{\phi}} 
\end{multline}
where, k is the Boltzmann’s constant and T is the temperature in Kelvin. Using equations (\ref{eq_pn}) and (\ref{eq_leeson}), phase noise of the oscillator in dBc/Hz can be given by the following expression.
\begin{equation}
\label{eq_pn_proposed}
\small
\mathcal{L}\{\Delta \omega\}  = 10\ log \bigg\{F\frac{4kTR_0^2}{R_m \times V_{OSC}^2}\ \bigg(\frac{\omega_0}{2Q_{OSC}\Delta\omega}\bigg)^2\bigg\} 
\end{equation}
Theoretical minimum phase noise of the proposed oscillator can be calculated by considering the noise due to $R_m$ only, for which $F=F_{min}=1$. 
Considering, $R_0 = R_m$ = 332 $\Omega$, $Q_{OSC}$ = 10K, $\omega_0$ = 30 GHz, $\Delta\omega$ = 1MHz, $V_{OSC} = V_{DD}=800 mV$, the theoretical minimum phase noise is $\mathcal{L}\{\Delta \omega\} $ = -167 dBc/Hz. This significantly low phase noise is a manifestation of effective utilization of high Q factor of the active-mode resonator. 
\section{Implementation and Simulation Results}
\label{impl_result}
\begin{figure}%
\centering
\includegraphics[scale=0.55]{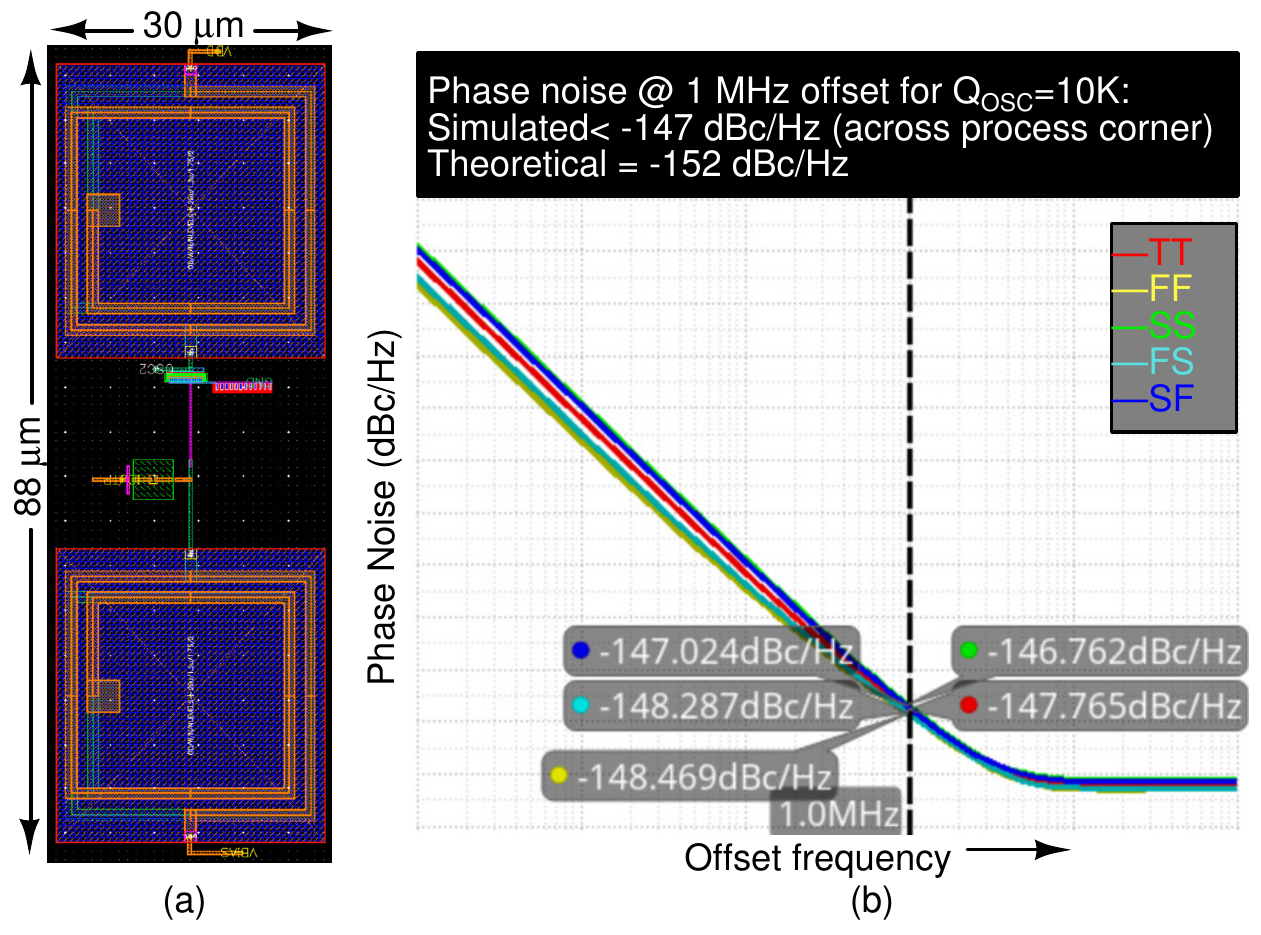}
\caption{(a) Layout of the proposed oscillator, (b) post-layout simulation results of PN across process corners showing close agreement with the theoretical model}
\label{fig_pn_process_tr}
\end{figure}

The proposed circuit (Fig. \ref{fig_ckt}(c)) has been implemented in 14 nm GF technology and post-layout simulations were performed on the design. Fig. \ref{fig_pn_process_tr}(a) shows the layout of the circuit, which occupies an area of 88$\mu$m $\times$ 30$\mu$m. 
In the design, %
$L_{0} = 650$ pH with $Q_{L_{0}} \approx 10$ and $R_{L_{0}} \approx 13 \Omega$, and $L_{\phi}=$ 650 pH, which satisfies equations (\ref{eq_l_phi_max}) and (\ref{eq_l_phi_min}). 
Moreover, $L_{\phi}$ with the extracted value of $C_{in} \approx 5$ fF, makes $f_{\phi} = \frac{1}{2\pi\sqrt{L_{\phi}C_{in}}}= 88$ GHz $>> f_0$ (30 GHz), 
and thus provides 90\textdegree\ phase shift with I-V conversion. 
For the designed $M_1$, $g_{m_{M1}} \approx 15$ mA/V. 
Electrical parameters of AM-RFT shown in Fig. \ref{fig_reso} were used for oscillator simulations. 
The total power consumption ($P_{DC}$) is 5.7 mW from 800 mV supply. 

\begin{figure}%
\centering
\includegraphics[scale=0.45]{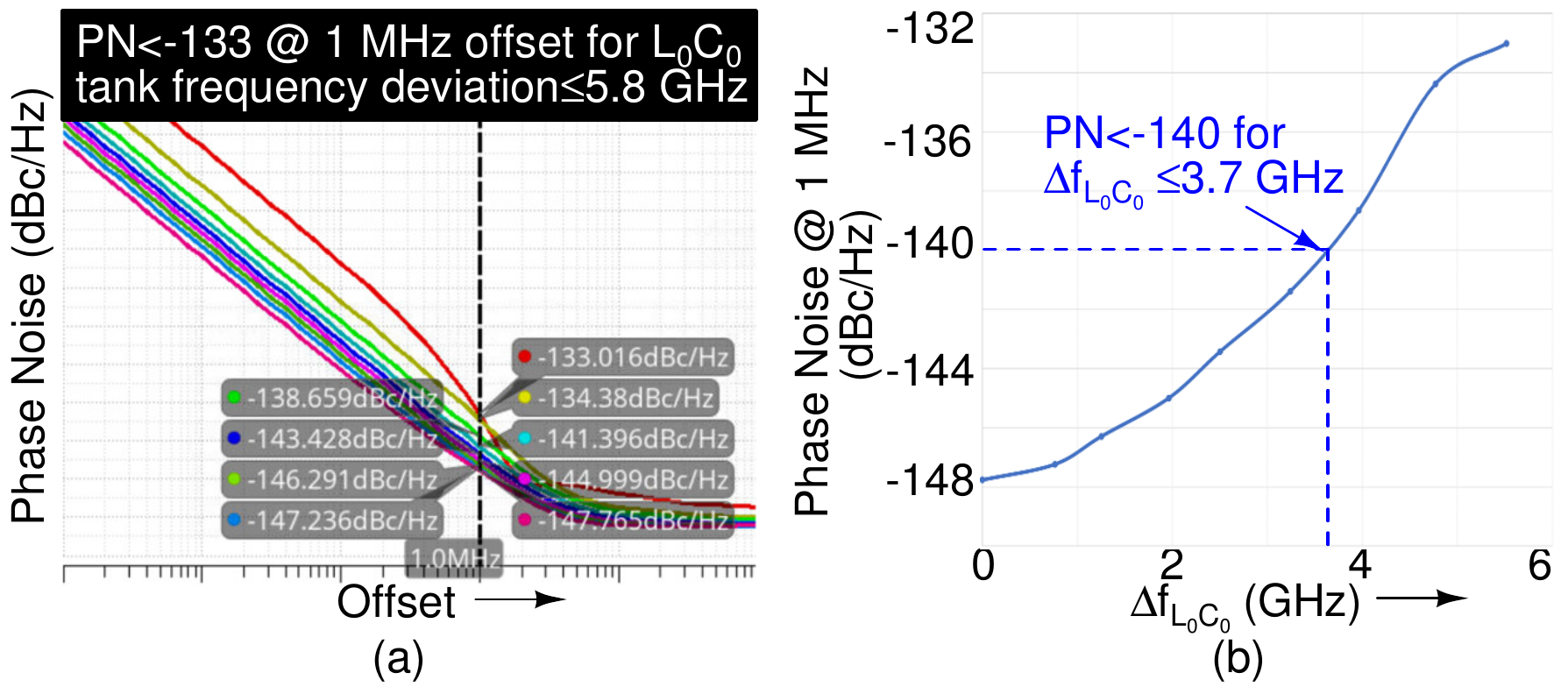}
\vspace{-2mm}
\caption{Post-layout simulation results showing (a) PN plots for different (than $f_0$) resonant frequencies of $L_0C_0$ tank 
(b) PN $<$ -140 dBc/Hz for $\Delta f_{L_0C_0} \leq 3.7$ GHz}
\label{fig_pn_f_deviation}
\vspace{-2mm}
\end{figure}

Fig. \ref{fig_pn_process_tr}(c) shows that the 
post-layout simulated phase noise at 1 MHz offset (PN)  %
is $<$-146 dBc/Hz across the process corners. %
With the values mentioned earlier and from Eq. (\ref{eq_noise_fac}), F=32.3, which gives a  
theoretical value of PN  %
$\approx$ -151.6 dBc/Hz (Eq. (\ref{eq_pn_proposed})), which is not too far from the simulated value of -147.8 dBc/Hz at TT as shown in Fig. \ref{fig_pn_process_tr}(b). 
FoM of the proposed oscillator is about 228 dBc/Hz. %
Fig. \ref{fig_pn_f_deviation}(a) shows PN plots 
when 
resonant frequency of $L_0C_0$ tank deviates with respect $f_0$. As depicted in Fig. \ref{fig_pn_f_deviation}(b), PN is better than -140 dBc/Hz for frequency deviations ($\Delta f_{L_0C_0}$) upto 3.7 GHz. 
This shows that the high-Q AM-RFT is able to sustain the oscillations by pulling the slight variations in $L_0C_0$ tank frequency to close to $f_0$. 
Table \ref{tab_results} shows the performance summary of the proposed oscillator and its comparison with other works,  
which conveys the fact that 
as compared to the oscillators with LC-tank resonators near 30 GHz, 
it is possible to utilize the high-Q of an active-mode monolithic MEMS resonator and build oscillators with significantly improved ($>$ 25 dB) PN and FoM. 

\begin{table}
\begin{center}
  \scriptsize
 \caption{Performance summary and comparison}
  \label{tab_results}
      \resizebox{1\linewidth}{!}{
  \begin{tabular}{|>{\centering}p{1.7cm}|>{\centering}p{1.06cm}|>{\centering}p{1.0cm}|>{\centering}p{1cm}|>{\centering}p{1.5cm}|>{\centering}p{1.4cm}|}
        \hline
         \textbf{Parameters} &\textbf{\cite{vco_17GHz_tcas2_2018}} & \textbf{\cite{vco_18_40GHz_isscc_2020}} & \textbf{\cite{vco_28GHz_IMS2020}} & \textbf{\cite{vco_28GHz_martin_isscc_2019}}  & \textbf{This work}\tabularnewline
        \hline
         Measured/ Simulated & Measured & Measured & Measured & Measured & Simulated (post-layout) \tabularnewline
        \hline
         Technology & $180$ nm CMOS & $40$ nm CMOS & $65$ nm CMOS & $65$ nm \\CMOS & $14$ nm CMOS\tabularnewline
         \hline        
        Frequency & 17 GHz & 30.7 GHz &28 GHz  & 29.92 GHz& 30 GHz\tabularnewline
        \hline
         Resonator & Switched-interdigital & Dual LC & Dual-path L \& C  & Multi-resonant RLCM & Active-Mode MEMS \cite{rft_isscc2018}  \tabularnewline
        \hline	
        PN (dBc/Hz) $@$1MHz & -105 & -104.98 & -102.5\dag  & -112.31 & $<$-140
        \tabularnewline
		\hline
		Power & 7.2 mW & $>$9 mW & 4.7 mW  & 4 mW & 5.7 mW  \tabularnewline
        \hline		
        FoM (dBc/Hz)\ddag & 182 & $<$184.4 & 184.5 &189.8 & 228\tabularnewline
        \hline          
         Supply & $1.8$ V & $1.1$ V &  $1.2$ V & $0.48$ V & $0.8$ V\tabularnewline
        \hline        
         Die-Area (mm$^2$)&  $0.34$ &  $0.08$ &  $0.0875$ & $0.08$ &  $0.0026$ \tabularnewline
        \hline                
  \end{tabular}
  }
\begin{tablenotes}
\item \dag Taken from the graph in the paper. 
\ddag$FoM = \big(\frac{f_0}{\Delta f}\big)^2/\big(\mathcal{L}\{\Delta f\} P_{DC}\big)\times 10^{-3}$
 \end{tablenotes}
\end{center}
\vspace{-8mm}
\end{table}
\vspace{-2mm}
\section*{Conclusions}
In this research a novel, low phase noise, monolithic oscillator topology has been presented with a high-Q active-mode MEMS resonator, which acts as a VCCS having 270\textdegree\ phase shift between its output current and input voltage. 
Supported with analysis, important insights have been presented in this brief, 
which indicate that 
the fundamental limits of PN and FoM for mmWave oscillators can be improved by $>$25 dB with  monolithic active-mode MEMS resonators as compared to the conventional LC tank resonators. 
The proposed topology is verified with the design and post layout simulations 
of the first 30 GHz oscillator with active mode resonator in 14-nm GF process. 
Post layout simulation results show that 
the proposed oscillator with active mode resonator having $Q=10K$, 
exhibits phase noise $<$ -140 dBc/Hz and FoM $>$ 228 dBc/Hz at 1 MHz offset for 30 GHz center frequency, which are $>$25 dB better as compared to the existing monolithic LC oscillators.  
\vspace{-2mm}
\addtolength{\textheight}{-12cm} %
\section*{Acknowledgement}
Authors acknowledge DARPA MIDAS program for funding this research work.
\bibliographystyle{IEEEtran}
\bibliography{ref_active_rft}
\end{document}